\documentclass[preprints,article,accept,moreauthors]{revtex4} 

\usepackage{graphicx}
\usepackage{epsfig}
\usepackage{wrapfig}
\usepackage{rotating}
\usepackage{color}
\usepackage{comment}
\usepackage[normalem]{ulem}
\usepackage{multirow}
\usepackage{tabularx}
\usepackage{graphicx}
\usepackage{float}
\usepackage{amsmath}
\usepackage{amssymb}
\usepackage{lipsum}
\usepackage{graphics}

\usepackage{hyperref}
\hypersetup{breaklinks=true,colorlinks=true,linkcolor=red,citecolor=blue,filecolor=cyan,urlcolor=magenta}

\usepackage{amsfonts}
\usepackage{pgfplots}
\usetikzlibrary{pgfplots.groupplots}
\usepackage{tikz}
\usetikzlibrary{positioning}
\usetikzlibrary{calc}
\usepackage{fp}
\usetikzlibrary{fixedpointarithmetic}
\usepackage[finalnew]{trackchanges}
\newlength\figureheight 
\newlength\figurewidth

\begin{document}

\title{Relativistic Magnetized Astrophysical Plasma Outflows in Black-Hole Microquasars}
\author{Theodora Papavasileiou$^{1}$}\email{th.papavasileiou@uowm.gr}
\author{Odysseas Kosmas$^{2}$}\email{odysseas.kosmas@manchester.ac.uk}
\author{Ioannis Sinatkas$^{1}$}\email{isinatkas@uowm.gr}
\affiliation{$^{1}$Department of Informatics, University of Western Macedonia, GR-52100 Kastoria, Greece}
\affiliation{$^{2}$Department of MACE, University of Manchester, George Begg Building, Manchester M1 3BB, UK} 

\begin{abstract}
Recently, microquasar jets have aroused the interest of many researchers focusing on the astrophysical plasma outflows and various jet ejections. In this work, we concentrate on the investigation of electromagnetic radiation and particle emissions from the jets of stellar black hole binary systems characterized by the hadronic content in their jets. Such emissions are reliably described within the context of the relativistic magneto-hydrodynamics. Our model calculations are based on the Fermi acceleration mechanism through which the primary particles (mainly protons) of the jet are accelerated. As a result, a small portion of thermal protons of the jet acquire relativistic energies, through shock-waves generated into the jet plasma. From the inelastic collisions of fast (non-thermal) protons with the thermal (cold) ones, secondary charged and neutral particles (pions, kaons, muons, $\eta$-particles, etc.) are created as well as electromagnetic radiation from the radio wavelength band, to X-rays and even to very high energy gamma-rays. One of our main goals is, through the appropriate solution of the transport equation and taking into account the various mechanisms that cause energy losses to the particles, to study the secondary particle concentrations within hadronic astrophysical jets. After testing our method on the Galactic MQs SS 433 and Cyg X-1, as a concrete extragalactic binary system, we examine the LMC X-1 located in the Large Magellanic Cloud, a satellite galaxy of our Milky Way Galaxy. It is worth mentioning that, for the companion O star (and its extended nebula structure) of the LMC X-1 system, new observations using spectroscopic data from VLT/UVES have been published few years ago.
\end{abstract}

\maketitle

\section{Introduction}\label{sec. 1}
 
The last decades, collimated plasma outflows have been observed to emerge from a wide variety 
of astrophysical objects. \add{These include} the proto-planetary nebulae, the compact objects (like galactic 
black holes or microquasars and X-ray binary stars), as well as the nuclei of active galaxies 
(AGNs) \cite{Heinz, Fabrika}. Despite their different scales concerning the size, the velocity, the amount of energy transported,
etc., these cosmic structures have strong similarities. The observations of multi-wavelength 
and multi-particle emissions from the black-hole X-ray binaries (BHXRBs) \cite{Aharonian, Albert2006, Albert2007} and the AGNs have shown that they are mainly due to the mass accretion onto the compact object (black hole or neutron star). \add{The accreted matter} \change{coming}{comes} from a companion (donor) 
star which is often a nearby main sequence star \cite{Falcke, Papav-Ody-Sinat-MDPI}. 
Thus, the inflow (from the companion star) into the accretion disc and the outflow in the jet 
motivate the investigation of the disc–jet connection \cite{Kording} in the latter systems. \add{They} \remove{which} are governed by 
non-linear partial differential equations that must be solved through the use of advanced numerical 
techniques \cite{Ody-Smpon-2015,Ody-Smpon-2017,Ody-Smpon-2018}.

Recently, the X-ray binary systems (XRBs) and the microquasars (MQ) along with their astrophysical 
magnetohydrodynamical outflows have aroused great interest among the researches dealing with the 
mechanisms of production and emission of high-energy particles (mostly neutrinos) and/or photons (X-rays,
gamma-rays, etc.) from such cosmic structures. \add{This is} not \add{restricted} only in the analysis of the \change{huge}{very large} amount of observations \cite{IceCube, Aharonian, Albert2006, Albert2007} \add{compared to emissions of smaller energy ranges} 
but also in their advanced theoretical modeling \cite{Romero_Review, Vieyro, Ody-Smpon-2015,Ody-Smpon-2017,Ody-Smpon-2018}. Up to now, multidimensional 
numerical simulations 
and theoretical calculations have attempted to shed light on the nature of the interaction mechanisms as well
as on the dynamics involved in the associated emissions \cite{Papav-Ody-Sinat-MDPI, Ody-Smpon-2018}. They 
offer good support for experimental efforts of astrophysical particles and radiation detection (for 
more details the reader is referred to Refs. \cite{Romero_Review, Vila}).
 
From an astronomy and astrophysics viewpoint, XRBs are binary systems consisted of a compact object (mostly stellar black hole or neutron star) and a 
donor star in rotational trajectory around the central star. Due to the presence of very high gravitational 
field around the compact object, mass is absorbed out of the companion star. \add{The result is} \remove{leading to} the creation of a 
rotating accretion disc of very high temperature matter and gas along the equatorial plane of the compact 
object. Subsequently, the magnetic field lines created by the rotating charged matter of the disc, collimate
the ejected plasma. \change{so as}{This way} two oppositely directed astrophysical jets are formed \cite{Mirabel}. From a magnetohydrodynamical
point of view, the jets (astrophysical plasma outflows) are considered as fluid flows emanating from the region of the compact object. Then, they may be strongly accelerated within a cone of radius $r(z)$ dependent on its half-opening angle $\xi$ ($r=ztan\xi$). \add{It is worth noting that, in some systems (e.g. the M87 and some AGNs) the consideration of a parabolic jet is geometrically more realistic. For example, in describing the jet acceleration, the parabolic shape favors the region near the jet's base. In the conical geometry, however, assumed in this work the acceleration region is put at greater distances from the base.}   

The plasma ejection is closely connected with the accretion disc formation and its
thickness as well as the jet creation mechanisms. Ideally, spherical symmetry would be well suited to describe thin accretion-disc plasma outflows. \remove{and,} Initially, prominent theoretical models have been developed by assuming isotropic emissions from the jets (in such models solution through analytic calculations are possible). Realistically (even under slow black-hole rotation and small-scale magnetic fields) there are rather strong deviations from a perfectly symmetrical geometry. In addition, the formation of two oppositely directed jets destroys the spherical symmetry of many systems which become mostly axisymmetric (around the $z$-axis, the jet-ejection axis). That is why in various types of microquasars (and AGNs) the statistical analysis was made on the basis of the jet's orientation. Moreover, astrophysical jets are often observed to be one-sided and associated with a Doppler factor that confirms the existence of bulk relativistic motion inside the jet. Recently, with the development of advanced efficient computational tools, the employment of more realistic, anisotropic emission 
(non-symmetric) models became possible. \add{A prominent example would be} \remove{like e.g.} the relativistic hydrocode 
PLUTO implemented in Refs. \cite{Ody-Smpon-2015,Ody-Smpon-2017,Ody-Smpon-2018,Beall}.      

For the purposes of our present study, we adopt a lepto-hadronic model for neutrino and gamma-ray production 
\cite{Romero, Reynoso, Bosch-Ramon}. \remove{in which} \add{Therefore, we consider that} the jet's matter consists mainly of hadrons and electrons (their portion is 
determined by defining the ratio $\alpha$ of protons to leptons) that are strongly collimated by the system's 
magnetic field. Furthermore, we assume that a portion of the main jet's content (electrons and protons) \cite{Begelman} is accelerated to rather 
relativistic velocities through shock-waves \cite{Berezhko, Gallant, Kirk, Achterberg}. \remove{in such a way that} \add{A power-law is best suited to describe} the energy distribution of
the fast protons $N'(E')$ \remove{follows a power-law}, which in the jet's rest frame, is given by the expression \\ 
\begin{equation}
\label{Eq1}
N'(E') = K_0 E'^{-2} \ GeV^{-1}cm^{-3} \, ,
\end{equation}
\\where $K_0$ is a normalization constant. The accelerated fast protons scatter with the cold protons of the
jet or the protons of the stellar wind \cite{Romero2003, Romero2007, Friend}. \change{or even}{They can also scatter} with the radiation fields (gamma-rays) emanating from sources inside or outside the jet. \change{and produce}{The result is} high-energy secondary particles (pions, muons, 
neutrinos, etc.) \add{production} as well as secondary gamma-ray photons \add{emission} through various reaction chains \cite{Reynoso, Romero, Kelner2006}.

In this work, initially, we discuss some of the possible interactions resulting in secondary particle and radiation 
production. In our previous works, we have addressed the proton-proton (p-p) mechanism and estimated the respective energy 
distributions as well as the emissivities of neutrino and gamma-ray emissions \cite{Papav-Ody-Sinat-MDPI, Papav-Papad-Kosm-JPCon, Papad-Papav-Kosm-JPCon}. 
In the \change{present work}{current study}, we concentrate on the proton-photon (p-$\gamma$) interaction mechanism and its dependence on geometric 
characteristics of the binary systems of our interest. \change{that}{These} include the Galactic Cygnus X-1 microquasar \cite{Ahnen} and the 
extragalactic LMC X-1 system \cite{Hyde} located at the neighboring galaxy of the Large Magellanic Cloud.  
For the detection of such emissions, extremely sensitive detector facilities, like the IceCube (at the South
pole), the ANTARES and KM3NeT (at the Mediterranean Sea), etc.,  are in operation to record the relevant signals reaching 
the Earth \cite{Aartsen2016, Adrian-Martinez}.

\section{ Radiation field density and transport equation in microquasar jets }\label{sec. 2}

The radiation fields that interact with the accelerated (fast) protons of the jet, may consist of soft X-ray photons 
emanating from the system's accretion disc. \add{They could also include} synchrotron radiation emitted by the charged particles (accelerated by the magnetic field inside the jet), or ultra violet (UV) photons originated from the corona 
region \cite{Cherepashchuk, Romero, Reynoso}. In the context of this study, we will take into account the first two cases that are involved in p-$\gamma$ collisions leading to a
reaction chain producing neutrinos and gamma-rays of high energies. We note that UV photons do not contribute significantly
to the energy range of our interest.

\subsection{Microquasar jet mechanisms leading to neutrino and gamma-ray production}

In our previous works \cite{Papav-Ody-Sinat-MDPI, Ody-Smpon-2015}, we have adopted the proton-proton (p-p) interaction mechanism taking place inside the relativistic astrophysical outflows of microquasar jets. \change{that leads}{This also leads} to the particle (pions, muons, neutrinos, etc.) production and radiation emission (gamma-rays, etc.) (see \cite{Papav-Ody-Sinat-MDPI}). Such emissions present axial symmetry around the jet's ejection $z$-axis. In Ref. \cite{Ody-Smpon-2017, Ody-Smpon-2018}, the employed PLUTO hydro-code permits emission calculations without the assumption of axial symmetry which are considered more realistic jet emission simulations. In this paper, however, we focus on the proton-photon (p-$\gamma$) interactions described below. 

The p-$\gamma$ mechanism reflects the collisions of the relativistic protons with the photons of the 
radiation fields discussed before. This results to the known photo-pion production shortly represented by the following scattering and/or decay reactions\\
\begin{align}
\label{gamma-interact} 
p + \gamma &\rightarrow p + \pi ^{0} \nonumber\\
p + \gamma &\rightarrow p + \pi ^{0} + \pi ^{+} + \pi^{-} \\
p + \gamma &\rightarrow n + \pi ^{+} \nonumber \, .
\end{align}
The above pions ($\pi ^0, \pi ^+, \pi ^-$) decay to gamma-ray photons, charged muons, neutrinos, etc., according to the following reactions 
\begin{align}
\label{pion-interact} 
\pi ^{0} &\rightarrow \gamma +\gamma \nonumber\\ 
\pi ^{+} &\rightarrow \mu ^{+} + \nu_{\mu}\rightarrow e^{+} + \bar{\nu }_{\mu} + \nu_{e} + \nu_{\mu} \\  
\pi^{-} &\rightarrow \mu ^{-} + \bar{\nu }_{\mu}\rightarrow e^{-} + \bar{\nu }_{e} + \nu_{\mu} + \bar{\nu }_{\mu} \nonumber \, .
\end{align} 
\\In the above reaction chains (\ref{gamma-interact} and \ref{pion-interact}), gamma-ray photons are emitted through the neutral 
pions decay. Also, through the subsequent decay of the secondary muons which create electrons $e^-$ 
(or positrons $e^+$), neutrinos or anti-neutrinos are produced too. By implementing the p-$\gamma$ mechanism, \remove{in the
relevant emissions investigated in our present study,} the above reactions constitute the main processes 
feeding the neutrino and gamma-ray production channels in the lepto-hadronic model employed \cite{Reynoso, Romero}. 

In general, the radiation density that interacts with the non-thermal (relativistic) protons, is due to mainly two contributing factors. The first is the synchrotron emission, by accelerated charged particles, that create a photon distribution, $n_{phS}$ (produced by relativistic electrons as well as protons). \add{It is} given by \cite{Reynoso, Romero} \\
\begin{equation}
\label{synchr_dens}
n_{phS}(\epsilon , z)
\approx \frac{\varepsilon _{syn}r(z)}{\epsilon c} \, ,
\end{equation}
\\ where $\varepsilon _{syn}$ corresponds to the total power radiated by electron and proton distributions and is given in Appendix [\ref{A1}]. The second contributing factor is an X-ray distribution, $n_{phX}$ (for $2 \ keV < \epsilon < 100 \ keV$), originated from the corona that surrounds the inner accretion disc. For the latter distribution, we have \cite{Cherepashchuk, Reynoso} \\
\begin{equation}
\label{x-ray_dens}
n_{phX}(\epsilon , z) = \frac{L_{X}e^{-\epsilon /kT_{e}}}{4\pi cz^{2}\epsilon ^{2}} \, ,  
\end{equation}
\\where X-ray luminosity is $L_{X}=10^{36}$ $ergs^{-1}$ and $kT_e\simeq 30$ $keV$.

Furthermore, within our approximation it holds $n_{ph}= n_{phX} + n_{phS}$. It should be noted that, the  primary particles (protons, electrons) as well as the secondary particles take part also in energy loss interactions and processes. \change{that}{These} include the energy losses due to jet adiabatic expansion, the losses due to inelastic collisions with the cold protons and losses due to the emission of synchrotron radiation \cite{Blumenthal, Papav-Papad-Kosm-JPCon, Papad-Papav-Kosm-JPCon}.  

Concerning the fast (non-thermal or relativistic) proton distribution, its shape in the one-zone approximation,
resembles to a power-law type (defined by a normalization constant $K_0$) \cite{Khangulyan} assumed in our model [see Eq. (\ref{Eq1})].

\subsection{ Solution of the transfer equation }

In the lepto-hadronic model employed in the present work, the acceleration mechanism (it is not included in the transport equation) is used to fix the injection function of the primary electrons and protons. \remove{and} Then, \change{to determine}{it determines} the maximum
energies that can \change{acquire}{be acquired by} the relativistic particles inside the jet. To this aim, the particle transfer (transport) equation 
must be solved which, assuming a steady-state model, is written as \cite{Ody-Smpon-2015,Ody-Smpon-2017,Ody-Smpon-2018}\\
\begin{equation}
\frac{\partial N(E,z)b(E,z)}{\partial E}+t^{-1}N(E,z)=Q(E,z) \, .
\label{tranf-equat}
\end{equation} 
\\ \remove{where}$N(E,z)$ represents the particle density per unit of energy ($cm^{-3}GeV^{-1}$) and $Q(E,z)$ denotes the particle source function (in units of $cm^{-3}GeV^{-1}s^{-1}$). Obviously, this is not a spherical but an axisymmetric model, hence the particle distributions and injection functions depend on $z$ (i.e, the distance to the central object on the ejection-axis of the jet). In the latter equation, $b(E)$ stands for the total energy loss 
rate given by\\
\begin{equation}
b(E)= \frac{dE}{dt}=-Et_{loss}^{-1} \, ,
\end{equation}
\\while $t^{-1}$ represents the particles' reduction rate as\\ 
\begin{equation}
t^{-1}=t_{esc}^{-1}+t_{dec}^{-1} \, .
\end{equation}
\\ \change{where}{Here,} $t_{dec}^{-1}$ is the decay rate (in the case of pions and muons) and $t_{esc}^{-1}$ the escape rate of the particles from the jet's region. The latter rate is given by\\
\begin{equation}
t_{esc}^{-1}=\bigg |\frac{c}{z_{max}-z_0}\bigg | \, , 
\end{equation}
\\with ($z_{max}-z_0$) being the length of the acceleration zone inside the jet.

The solution of the differential equation (\ref{tranf-equat}) is written as\\
\begin{equation}
N(E,z)=\frac{1}{\mid b(E) \mid}\int_{E}^{E_{max}} Q(E',z)e^{-\tau (E,E')}dE' \, ,
\end{equation}
\\where the maximum proton energy is approximated by $E_{p}^{max}\simeq 10^7$ $GeV$, while\\
\begin{equation}
\tau (E,E')=\int_{E}^{E'} \frac{dE"t^{-1}}{\mid b(E")\mid} \, .
\end{equation}
\\The source function $Q(E,z)$, for the relativistic particles (in the jet's rest frame) is given by \cite{Reynoso}\\
\begin{align}
Q(E',z)=Q_0\left(\frac{z_0}{z}\right)^3E'^{-2} \, .
\label{Prot_sourc}
\end{align}
\\
In the observer's reference frame, $Q(E,z)$ takes the form given in the Appendix [see Eq. (\ref{eq6}) of \ref{A2}] where the normalization constant $Q_0$ of Eq. (\ref{Prot_sourc}) is also given in the Appendix.  


\begin{table}[H]
\caption{\label{Table1} Model parameters describing geometric characteristics of the Galactic Cygnus X-1 binary system and the extragalactic 
LMC X-1 located at the Large Magellanic Cloud.}
\begin{center}
\begin{tabular}{l l l l}
\hline 
\textbf{Description} & \textbf{Parameter} & \textbf{Cygnus X-1} & \textbf{LMC X-1} \\
\hline
Jet's base & $z_0$ ($cm$) & $191R_{Sch}$ & $95R_{Sch}$ \\ [0.2ex]
Acceleration zone limit & $z_{max}$ ($cm$) & $956R_{Sch}$ & $477R_{Sch}$ \\ [0.2ex]
Mass of compact object & $M_{BH}$ & 14.8\(M_\odot\)\cite{Orosz2011} & 10.91\(M_\odot\)\cite{Orosz2009} \\ [0.2ex]
Angle to the line-of-sight & $\theta$ ($^\circ$) & 27.1\cite{Orosz2011} & 36.38\cite{Orosz2009} \\ [0.2ex]
Jet's half-opening angle & $\xi$ ($^\circ$) & 1.5\cite{Stirling} & 3\footnotemark[1] \\ [0.2ex]
Jet's bulk velocity & $\upsilon _{b}$ & 0.6c\cite{Stirling} & 0.92c\footnotemark[1] \\ [0.2ex]
\hline
\end{tabular}
\end{center}
\end{table} 
\footnotetext[1]{indicative values that we consider for our calculations.}

\section{Interaction frequency and particle emission through p-$\gamma$ mechanism}\label{sec. 3}

The injection function for the pions produced through the p-$\gamma$ interaction 
mechanism is the following \\
\begin{equation}
\label{pg-int}
\begin{split}
Q_{\pi}^{(p\gamma)}(E,z) &=5N_p(5E,z)\omega_{p\gamma }(5E, z)\bar{N}_{\pi }^{(p\gamma )}(5E) \, ,
\end{split}
\end{equation}
\\where $N_p(E,z)$ denotes the proton energy distribution. \change{and}{Also,} $\omega_{p\gamma}$ is the p-$\gamma$ collision frequency that results in pion production. \remove{Also,} The mean number $\bar{N}_{\pi}^{(p\gamma )}$ of positive or negative pions produced per p-$\gamma$ collision is given by \\ 
\begin{equation}
\bar{N}_{\pi}^{(p\gamma )}=p_{p\rightarrow n}p_1+2p_2 \, ,
\end{equation}
\\where the parameter $p_{p\rightarrow n}\simeq 0.5$ is to express the probability of conversion of a primary proton to a neutron. Also, $p_1$ and $p_2$ are defined as \\ 
\begin{equation}
p_1=\frac{K_2-\bar{K}_{p\gamma}}{K_2-K_1} \, , 
\end{equation}
\\ and $p_2=1-p_1$. Moreover, we have $K_1=0.2$ and $K_2=0.6$. In the latter equation, for the mean inelasticity parameter $\bar{K}_{p\gamma}$ it holds \\
\begin{equation}
\label{Kpg}
\bar{K}_{p\gamma}=\frac{t_{p\gamma}^{-1}}{\omega _{p\gamma}} \, .
\end{equation}
\\The p-$\gamma$ collision frequency $\omega_{p\gamma}$ in Eq. (\ref{pg-int}) and (\ref{Kpg}) is given by \cite{Atoyan, Reynoso} \\
\begin{equation}
\omega_{p\gamma}(E_p,z)=\frac{c}{2\gamma_p^{2}}\int_{\epsilon _{th}/2\gamma_p}^{\infty}\frac{n_{ph}(\epsilon , z)}{\epsilon ^2}d\epsilon\int_{\epsilon _{th}}^{2\epsilon\gamma _p}\sigma _{p\gamma}^{(\pi )}(\epsilon ')\epsilon 'd\epsilon ' \, ,
\end{equation}
\\where $\gamma _p=E_p/m_pc^2$, $n_{ph}(\epsilon , z)$ is the radiation field density that was discussed in sec. \ref{sec. 2}. \remove{and} The threshold energy $\epsilon _{th}$ is assumed to be $\epsilon _{th}=0.15$ $GeV$ \cite{Atoyan}.

After the above definitions, the respective cross-section $\sigma_{p\gamma}^{(\pi)}$ is given by \cite{Atoyan, Kelner} \\
\begin{equation}
\sigma_{p\gamma}^{(\pi)}=[3.4\Theta (\epsilon '-0.2)\Theta (0.5-\epsilon ')+1.2\Theta (\epsilon '-0.5)]\times 10^{-28}cm^2 \, ,
\end{equation}
\\where $\Theta (x)$ denotes the well-known step function. In the latter relationships, the proton-photon collision rate $t_{p\gamma}^{-1}$ was obtained from the expression \cite{Atoyan}\\
\begin{equation}
\label{Eq18}
t_{p\gamma}^{-1}=\frac{c}{2\gamma _p^2}\int_{\epsilon _{th}/2\gamma_p}^{\infty}\frac{n_{ph}(\epsilon , z)}{\epsilon ^2}d\epsilon\int_{\epsilon _{th}}^{2\epsilon\gamma _p}\sigma _{p\gamma}^{(\pi )}(\epsilon ')K _{p\gamma}^{(\pi )}(\epsilon ')\epsilon 'd\epsilon ' \, .
\end{equation}
\\ \change{where}{In the above expression,} $K _{p\gamma}^{(\pi )}(\epsilon ')$ is the respective inelasticity function \cite{Atoyan, Kelner} which is equal to \\
\begin{equation}
K _{p\gamma}^{(\pi )}(\epsilon ')=0.2\Theta (\epsilon '-0.2)\Theta (0.5-\epsilon ')+0.6\Theta (\epsilon '-0.5) \, .
\end{equation}
\\
\section{Results and discussion}\label{sec. 4}

\begin{figure}[ht] 
\begin{tikzpicture}
\centering
  \node (img1)  {\includegraphics[width=.523\linewidth]{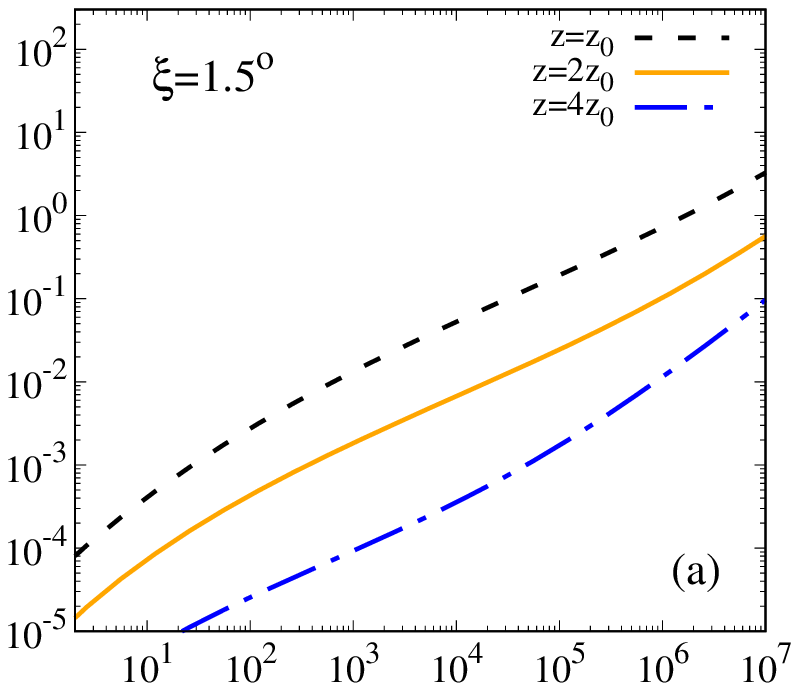}};
  \node[left= of img1, node distance=0cm, rotate=90, anchor=center,yshift=-1.2cm, font=\color{black}, font=\large] {$t_{p\gamma}^{-1} [s^{-1}]$};
  \node[right= of img1, xshift=-1.8cm] (img2)  {\includegraphics[width=.523\linewidth]{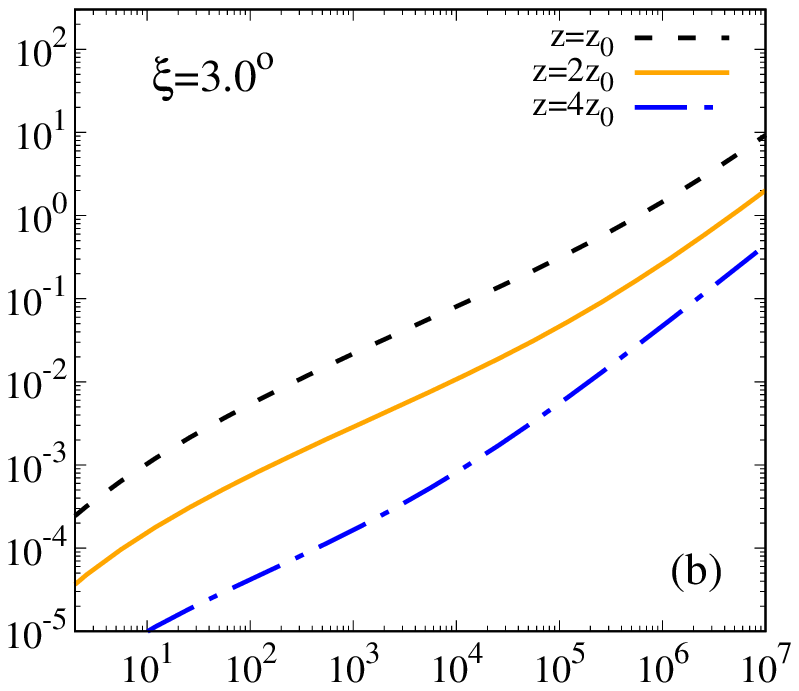}};
  \node[below= of img1,yshift=1cm] (img3)  {\includegraphics[width=.523\linewidth]{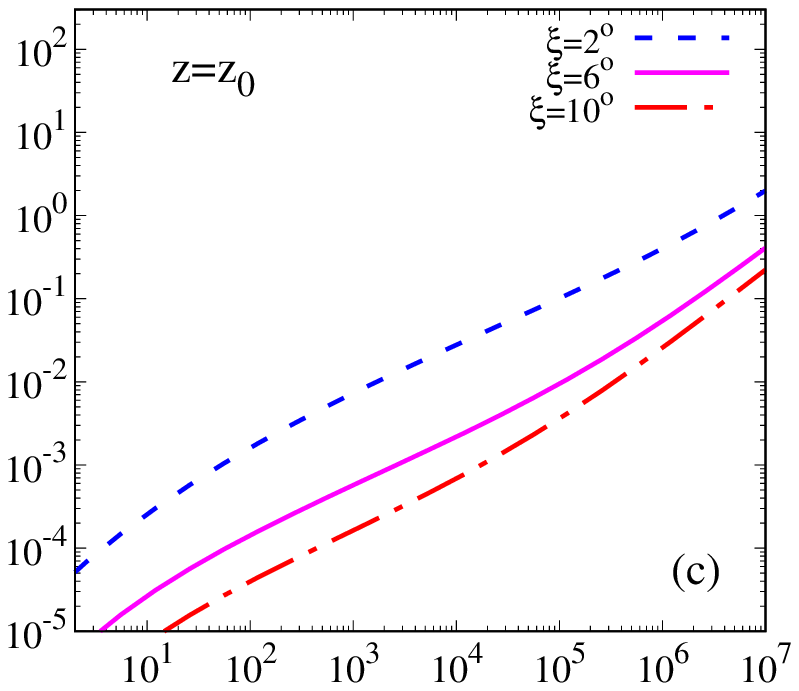}};
  \node[below= of img3, node distance=0cm, yshift=1cm, xshift=0.5cm, font=\color{black}, font=\large] {E [GeV]};
  \node[left= of img3, node distance=0cm, rotate=90, anchor=center,yshift=-1.2cm, font=\color{black}, font=\large] {$t_{p\gamma}^{-1} [s^{-1}]$};
  \node[right= of img3, xshift=-1.8cm] (img4)  {\includegraphics[width=.523\linewidth]{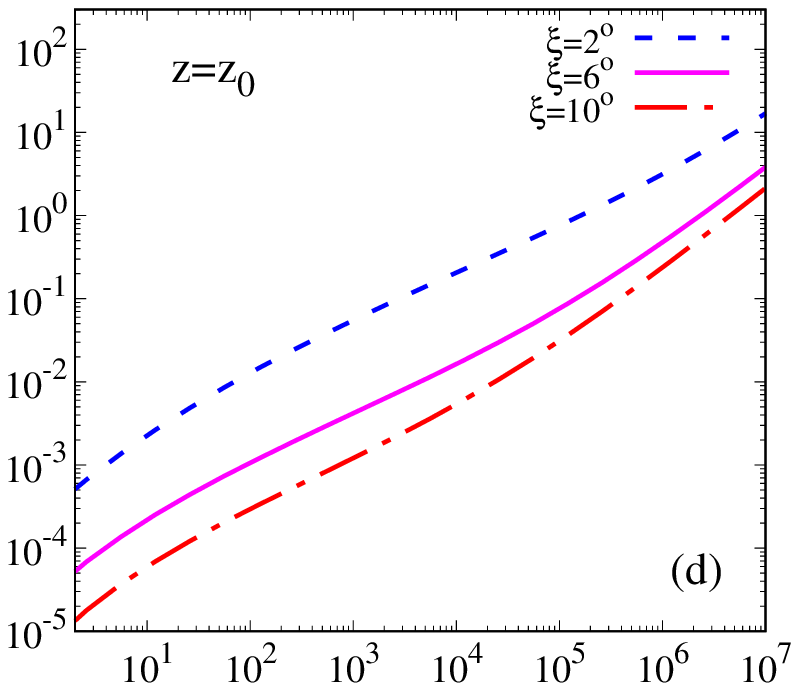}};
  \node[below= of img4, node distance=0cm, yshift=1cm, xshift=0.5cm, font=\color{black}, font=\large] {E [GeV]};
\end{tikzpicture}

\caption{\label{figure2} Proton-photon (p-$\gamma$) interaction rate $t_{p\gamma}^{-1}$ as a function of proton energy E for the binary systems Cygnus X-1 (left column) and LMC X-1 (right column). In the upper two sub-figures (\textbf{a}, \textbf{b}), we show the p-$\gamma$ interaction rate for three different values of the distance z (in factors of $z_0$) inside the jet from the compact object. In the lower two sub-figures (\textbf{c}, \textbf{d}), we plot the $t_{p\gamma}^{-1}$ for three different values of the jet's half-opening angle $\xi$.}
\end{figure}
In the context of the model chosen in this study, at first the p-$\gamma$ collision rate $t_{p\gamma}^{-1}$ of Eq. (\ref{Eq18}) was calculated for the X-ray binary systems Cygnus X-1 and LMC X-1. The geometric properties and other parameters of these systems are listed in Table \ref{Table1}. The results obtained for the $t_{p\gamma}^{-1}$ in the latter microquasars are presented in Fig. \ref{figure2}. In the upper two sub-figures, we illustrate the interaction rate as a function of the proton energy E for three different distances of the studied point up to the system's central object (z). The range of $z$ covers the length of the acceleration zone (from $z_0$ to $z_{max}=5z_0$). It is \change{obvious}{supported by theoretical calculations} that, for greater distances, the p-$\gamma$ collision rate decreases. The reason is the dependence of X-ray photon density [see Eq. (\ref{x-ray_dens})] as well as the fast particle density on $z$. It is reasonable that as the jet expands (e.g., its radius increases), the mean-free path of the particles and photons involved in the collisions increases as well. That leads to reduced collision rates. 

In the lower two sub-figures of Fig. \ref{figure2}, we illustrate the p-$\gamma$ collision rate for three different values of the half-opening angle $\xi$ of the jet. \change{that}{This angle} strongly depends on the magnetic field strength and characterizes the jet's collimation. The values of $\xi$ considered are representative and cover the assumed range extended up to $10^o$. We notice that the collision rate decreases as the jet expands perpendicularly to its ejection axis (i.e., for greater angles). \remove{Of course,} This is \change{expected}{validated by the corresponding theoretical expressions} if we consider the decrease of the synchrotron emission which is due to the magnetic field that controls the jet's collimation. In addition, the mean-free path of the proton-photon collisions increases causing the reduction of the respective rate. 

\begin{figure}[ht]
\begin{tikzpicture} 
\centering
\node (Cpth) {\includegraphics[width=.523\linewidth]{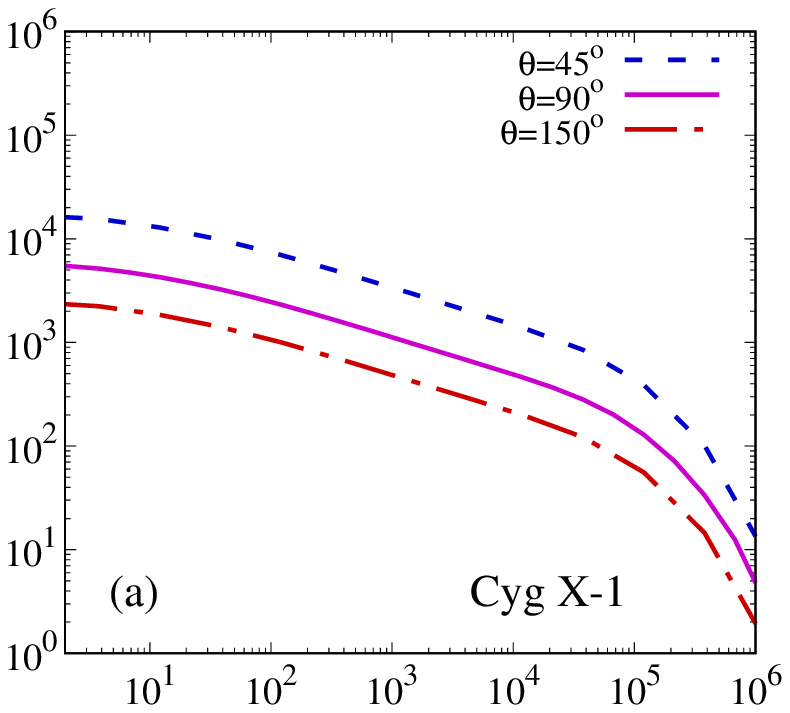}};
\node[below= of Cpth, node distance=0cm, yshift=1cm, xshift=0.5cm, font=\color{black}, font=\large] {E [GeV]};
\node[left= of Cpth, node distance=0cm, rotate=90, anchor=center, yshift=-1.2cm, font=\color{black}, font=\large] {$N_{\pi}^{p\gamma} [GeV^{-1}cm^{-3}$]};
\node[right= of Cpth, xshift=-1.8cm] (Cpub) {\includegraphics[width=.523\linewidth]{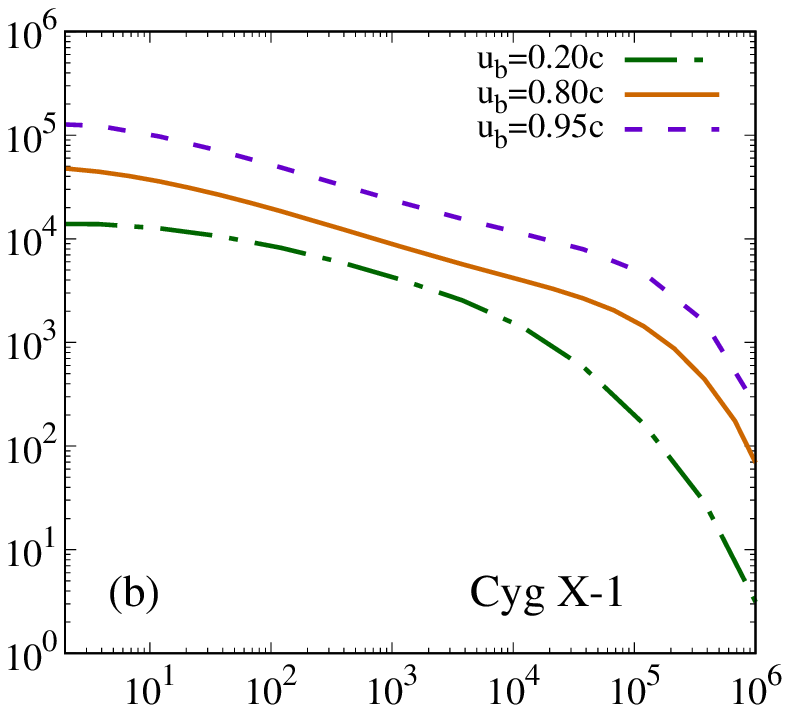}};
\node[below= of Cpub, node distance=0cm, yshift=1cm, xshift=0.5cm, font=\color{black}, font=\large] {E [GeV]};
\end{tikzpicture}

\caption{\label{figure3} For Cygnus X-1, we present pion distributions for different values of the system's angle to the line of sight (\textbf{a}) and bulk velocity of the jet's ejected matter (\textbf{b}). The pions in this case have not been subjected to energy loss mechanisms.}
\end{figure}

We have, also, calculated the pion energy distributions for different typical values of the angle ($\theta$) between the jet axis and the direction of the line of sight as well as the bulk velocity of the jet's matter ($u_b$) for the Cygnus X-1 binary system, see Fig. \ref{figure3}. It is evident that the particle production is higher for smaller values of $\theta$ as demonstrated in Eq. (\ref{eq6}) in the Appendix [\ref{A2}]. Also, the particle production increases for greater bulk velocities $u_b$ as shown in sub-figure (b) of Fig. \ref{figure3}. \change{This is rather an obvious result}{This result is justified by} considering the relation of bulk velocity to the jet's collimation and the proton-photon interaction rate which increases for larger energies. \add{It is hoped that, our present results would be helpful for future relevant experimental and observational measurements.}

\begin{figure}[ht]
\begin{tikzpicture} 
\centering
\node (Cpa) {\includegraphics[width=.523\linewidth]{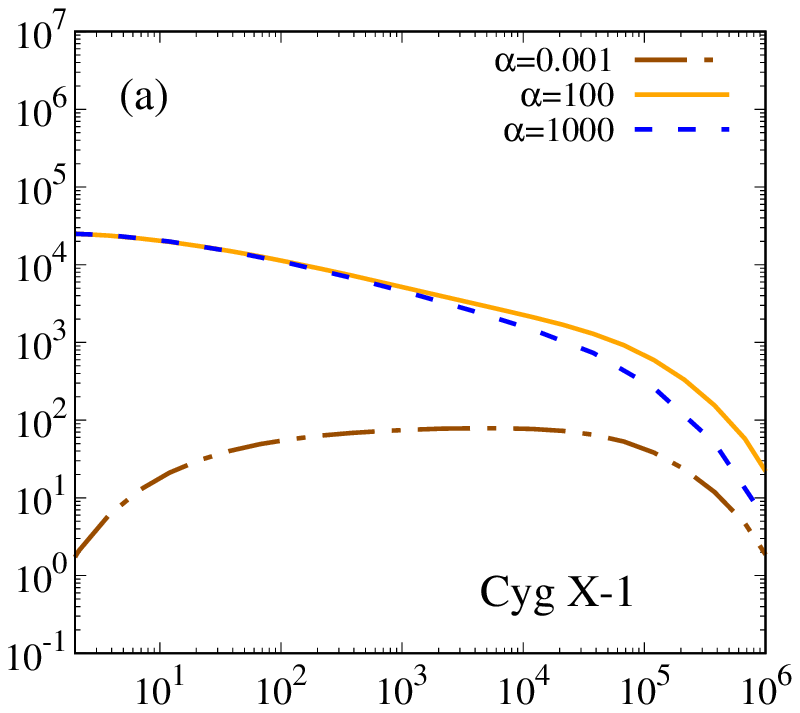}};
\node[left= of Cpa, node distance=0cm, rotate=90, anchor=center, yshift=-1.2cm, font=\color{black}, font=\large] {$N_{\pi}^{p\gamma} [GeV^{-1}cm^{-3}$]};
\node[below= of Cpa, node distance=0cm, yshift=1cm, xshift=0.5cm, font=\color{black}, font=\large] {E [GeV]};
\node[right= of Cpa, xshift=-1.8cm] (Lpa) {\includegraphics[width=.523\linewidth]{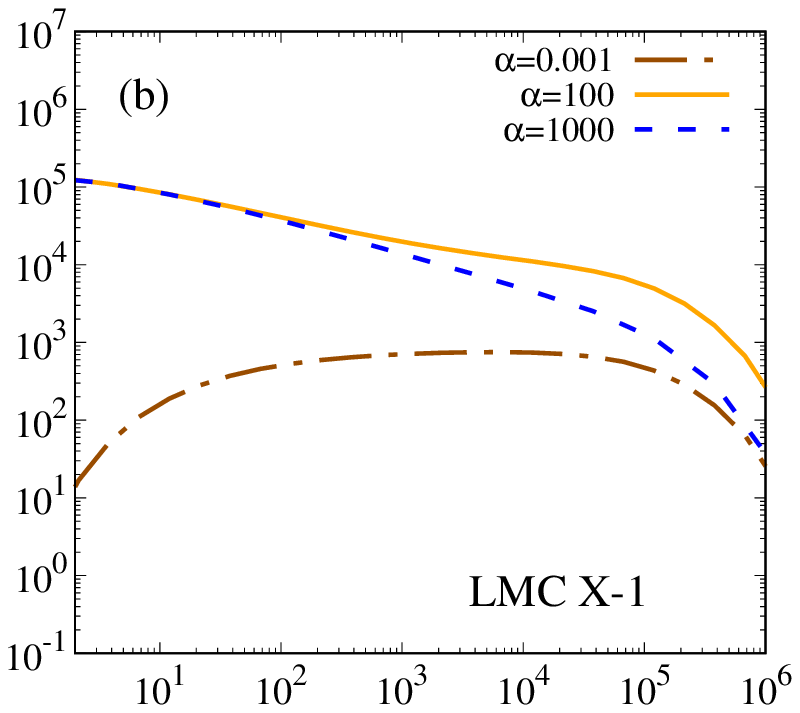}};
\node[below= of Lpa, node distance=0cm, yshift=1cm, xshift=0.5cm, font=\color{black}, font=\large] {E [GeV]};
\end{tikzpicture}

\caption{\label{figure4} Pion energy distributions from p-$\gamma$ collisions in the jets of binary systems Cygnus X-1 (\textbf{a}) and LMC X-1 (\textbf{b}) for three different values of hadron-to-lepton ratio $\alpha$. }
\end{figure}

In this work, we, furthermore study the pion distributions produced by p-$\gamma$ interactions for three different values of the parameter $\alpha$ defined as \\
\begin{equation}
\alpha =\frac{L_h}{L_e} \, .
\end{equation}  
\\$L_h$ and $L_e$ denote the hadronic and leptonic luminosity respectively. These values correspond to: (i) a leptonic model ($\alpha =0.001$), (ii) a hadronic model ($\alpha =100$) and (iii) an extreme-hadronic model ($\alpha =1000$). The results are shown in Fig. \ref{figure4}. Our main purpose, here, is the comparison between the particle (pions) production in these well-implemented models (leptonic and hadronic). \remove{and} For that reason, we have not considered the possible energy losses that the aforementioned particles are subjected through the various mechanisms \cite{Papav-Papad-Kosm-JPCon, Papad-Papav-Kosm-JPCon}. In Fig. \ref{figure4}, we notice the reasonable decrease in pion production in the leptonic case as the protons are reduced compared to electrons inside the jet. However, as can be seen, there is no essential difference between the hadronic and the extreme-hadronic model, therefore, a ratio of $\alpha\simeq 100$ is deemed sufficient enough. 
      
\section{Summary and Conclusions}\label{sec. 5}

Astrophysical binary systems, consisted of a high-mass compact object (Black Hole or Neutron star) 
that absorbs mass out of a companion star (forming a rotating accretion disc) are studied extensively in recent years. More specifically, their magnetohydrodynamical astrophysical flow ejection and high-energy radiation as well as particle emissions have been the research subject of many authors working in this field. In our present work, we go beyond the spherically symmetric models of neutrino and gamma-ray emissions and adopt \change{conic}{conical} jets, i.e. axially symmetric plasma outflows from microquasar jets. 

We, furthermore, assume that strong shock-waves accelerate a portion of the jets' charged particles (mostly protons but also electrons) to rather relativistic energies obeying a power-law energy-dependent distribution. Under the above circumstances, these particles interact with thermal protons of the jet (or of stellar winds) or with the radiation fields originating from energy-exchanging mechanisms (i.e. between electrons and low-energy photons) inside or outside the jet's region. The outcomes of these interactions are secondary particles and photons such as pions, muons, neutrinos, gamma-ray photons, etc. The multi-messenger signals created this way are detectable by the terrestrial extremely sensitive detectors like the IceCube, ANTARES, KM3NeT, etc.       

In our study, we mainly focus on the p-$\gamma$ interaction mechanism (which was not taken explicitly into account in our previous works) and the parameters (e.g., geometric characteristics of the binary systems, the jet's matter composition of leptons or hadrons, etc.) which affect the associated emissions of particle and radiation. In particular, we have selected two concrete examples, the Galactic Cygnus X-1 and the extragalactic LMC X-1 binary systems, to study with the model chosen
and present numerical calculations for the collision rates and energy distributions of particles that come out (pions). Our results show that the particle density production strongly increases in more collimated flow ejections which is highly dependent on the prevailing magnetic field. In addition, we find that the smaller the distance z from the central region the larger the production of particles and radiation is emitted.

\vspace{6pt} 

\appendix
\section{}
\subsection{Synchrotron power radiation by particle distributions}
\label{A1}

When a charged particle of energy $E$ is being accelerated by a magnetic field with pitch angle $a$, it emits synchrotron radiation. The power of the radiation per units of the emitted photons' energy and pitch angle $a$ is given by \cite{Blumenthal}\\
\begin{equation}
\label{Psyn}
P_{syn}(\epsilon , E, z, \alpha)=\frac{\sqrt{3}e^3B(z)sin(a)}{hmc^2}\frac{\epsilon}{E_{cr}}\int_{\frac{\epsilon}{E_{cr}}}^{\infty} K_{5/3}(\zeta )d\zeta \, ,
\end{equation}
\\where $B(z)$ is the binary system's magnetic field responsible for the jet's collimation. $e$ is the electric charge, $h$ denotes the Planck constant and $E_{cr}$ corresponds to the critical frequency of the emitted radiation which is written as\\
\begin{equation}
E_{cr}=\frac{3heB(z)sin(a)}{4\pi mc}\gamma ^2 \, . 
\end{equation} 
\\In Eq. (\ref{Psyn}), we integrate over the modified Bessel function of order 5/3 which we calculate through the following relationships\\
\begin{align}
K_{\frac{1}{3}}(\zeta )&=\sqrt{3}\int_{0}^{\infty} exp\left[-\zeta\left(1+\frac{4x^2}{3}\right)\sqrt{1+\frac{x^2}{3}}\right]dx \\
K_{\frac{2}{3}}(\zeta )&=\frac{1}{\sqrt{3}}\int_{0}^{\infty}\frac{3+2x^2}{\sqrt{1+\frac{x^2}{3}}}exp\left[-\zeta \left(1+\frac{4x^2}{3}\right)\sqrt{1+\frac{x^2}{3}}\right]dx \\
K_{\frac{5}{3}}(\zeta)&=K_{\frac{1}{3}}(\zeta)+\frac{4}{3\zeta}K_{\frac{2}{3}}(\zeta) \, .
\end{align}
\\After calculating $P_{syn}$, we integrate over the pitch angle $a$ as well as the particle energy distribution in order to obtain the total power radiated per unit energy by a particle (electron or proton) distribution such as those we discussed before. The result for electron or proton distributions is given below\\  
\begin{equation}
\varepsilon _{syn}^{(e,p)}(\epsilon )=\int d\Omega _{\alpha}\int_{E_{e,p}^{(min)}}^{E_{e,p}^{(max)}} P_{syn}N_{e,p}(E,z)dE \, ,
\end{equation}
\\where $E_{p}^{min}=1.2$ $GeV$ and $E_{e}^{min}=0.001$ $GeV$ are the minimum proton and electron energies respectively. For the maximum energies, we have $E_{p}^{max}\simeq 10^7$ $GeV$ and $E_{e}^{max}\simeq 7$ $GeV$.

The total power radiated by both electrons and protons is given by\\
\begin{equation}
\varepsilon _{syn}(\epsilon )= \varepsilon _{syn}^{(e)}(\epsilon ) + \varepsilon _{syn}^{(p)}(\epsilon )\, .
\end{equation}
\\  

\subsection{Injection function in observer's reference frame}
\label{A2}

The transformation of the proton injection function to the observer's reference frame is given by \cite{Torres, Ody-Smpon-2017, Ody-Smpon-2018}\\
\begin{align}
\label{eq6}
Q(E,z) = Q_0\left(\frac{z_0}{z}\right)^3\frac{\Gamma_b ^{-1}(E - \beta_b cos\theta\sqrt{E^2-m^2c^4})^{-2}}
{\sqrt{sin^2\theta + \Gamma_b^2\left(cos\theta -\frac{\beta_bE}{\sqrt{E^2-m^2c^4}}\right)^2}} \, .
\end{align}
\\ \change{where}{Here,} $\Gamma _b$ is the Lorentz factor responding to the jet's bulk velocity ($u_b=\beta _bc$ and $\Gamma _b=(1-\beta _b^2)^{ -\frac{1}{2}}$) and $Q_0$ is given by \\
\begin{align}
Q_0=\frac{8q_rL_k}{z_0r_0^2ln(E_p^{max}/E_p^{min})} \, ,
\end{align}
\\where $r_0$ is the jet radius that corresponds to the distance $z_0$ from the central object. Moreover, the kinetic luminosity that is transferred in the jet $L_k$ is considered to be 10\% of the central object's Eddington luminosity \cite{Kording}. We, also, adopt the value $q_r=0.1$ for the portion of relativistic protons and electrons inside the jet.  

\section*{Data Availability}

There is no data used in this paper.

\section*{Conflicts of Interest} The authors declare that there are no conflicts of interest regarding the publication of this paper.

\section*{Acknowledgments} This research is financed in the context of the PhD dissertation of T.P by the Department of Informatics, School of Sciences of University of Western Macedonia. 


\begin{thebibliography}{99}
\bibitem[(2002)] {Heinz} Heinz, S.; Sunyaev, R. Cosmic rays from microquasars: A narrow component to the CR spectrum? {\em Astron. Astrophys.} {\bf 2002}, {\em 390}, 751--766.
\bibitem[(2004)] {Fabrika} Fabrika, S. The jets and supercritical accretion disk in SS433. {\em Astrophys. Space Phys. Rev.} {\bf 2004}, {\em 12}, 1--152.
\bibitem[(2005)] {Aharonian} Aharonian, F.; Akhperjanian, A.G.; Aye, K.M.; Bazer-Bachi, A.R.; Beilicke, M.; Benbow, W.; Berge, D.; Berghaus, P.; Bernlöhr, K.; Boisson, C.; et al. Discovery of Very High Energy Gamma Rays Associated with an X-ray Binary. {\em Science} {\bf 2005}, {\em 309}, 746--749.
\bibitem[(2006)] {Albert2006} Albert, J.; Aliu, E.; Anderhub, H.; Antoranz, P.; Armada, A.; Asensio, M.; Baixeras, C.; Barrio, J.A.; Bartelt, M.; Bartko, H.; et al. Variable Very-High-Energy Gamma-Ray Emission from the Microquasar LS I +61 303. {\em Science} {\bf 2006}, {\em 312}, 1771--1773.
\bibitem[(2007)] {Albert2007} Albert, J.; Aliu, E.; Anderhub, H.; Antoranz, P.; Armada, A.; Baixeras, C.; Barrio, J.A.; Bartko, H.; Bastieri, D.; Becker, J.K.; et al. Very High Energy Gamma-Ray Radiation from the Stellar Mass Black Hole Binary Cygnus X–1. {\em Astrophys. J.} {\bf 2007}, {\em 665}, L51--L54.
\bibitem[(1994)] {Falcke} Falcke, H.; Biermann, P. The jet-disk symbiosis. 1. Radio to X-ray emission models for quasars. {\em Astron. Astrophys.} {\bf 1994}, {\em 293}, 665--682.
\bibitem[(2021)] {Papav-Ody-Sinat-MDPI} Papavasileiou, Th.V.; Kosmas, O.T.; Sinatkas, J. Simulations of Neutrino and Gamma-Ray Production from Relativistic Black-Hole Microquasar Jets. {\em Galaxies} {\bf 2021}, {\em 9(3)}, 67.
\bibitem[(2006)] {Kording} Körding, E.G.; Fender, R.P.; Migliari, S. Jet-dominated advective systems: Radio and X-ray luminosity dependence on the accretion rate. {\em Mon. Not. R. Astron. Soc.} {\bf 2006}, {\em 369}, 1451--1458.
\bibitem[(2015)] {Ody-Smpon-2015} Smponias, T.; Kosmas, O. High Energy Neutrino Emission from Astrophysical Jets in the Galaxy. {\em Adv. High Energy Phys.} {\bf 2015}, 921757.
\bibitem[(2017)] {Ody-Smpon-2017} Smponias, T.; Kosmas, O.  Neutrino Emission from Magnetized Microquasar Jets. {\em Adv. High Energy Phys.} {\bf 2017}, 496274.
\bibitem[(2018)] {Ody-Smpon-2018} Kosmas, O.T.; Smponias, T.  Simulations of Gamma-Ray Emission from Magnetized Microquasar Jets. {\em Adv. High Energy Phys.} {\bf 2018}, 960296. 
\bibitem[(2018)] {IceCube} Aartsen, M.; Ackermann, M.; Adams, J.; Aguilar, J.A.; Ahlers, M.; Ahrens, M.; Al Samarai, I.; Altmann, D.; Andeen, K.; Anderson, T.; et al. Neutrino emission from the direction of the blazar TXS 0506+056 prior to the IceCube-170922A alert. {\em Science} {\bf 2018}, {\em 361}, 147--151.
\bibitem[(2017)] {Romero_Review} Romero, G.E.; Boettcher, M.; Markoff, S.; Tavecchio, F. Relativistic Jets in Active Galactic Nuclei and Microquasars. {\em Space Sci. Rev.} {\bf 2017}, {\em 207}, 5--61.
\bibitem[(2012)] {Vieyro} Vieyro, F.L.; Romero, G.E. Particle transport in magnetized media around black holes and associated radiation. {\em Astron. Astrophys.} {\bf 2012}, {\em 542}, A7.
\bibitem[(2008)] {Vila} Romero, G.E.; Vila, G.S. The proton low-mass microquasar: High-energy emission. {\em Astron. Astrophys.} {\bf 2008}, {\em 485}, 623–631.
\bibitem[(1999)] {Mirabel} Mirabel, I.F.; Rodríguez, L.F. Sources of Relativistic Jets in the Galaxy. {\em Annu. Rev. Astron. Astrophys.} {\bf 1999}, {\em 37}, 409--443.
\bibitem[(2018)] {Beall} Beall, J.H. A Review of Astrophysical Jets. {\em PoS Proc. Sci.} {\bf 2018}, {\em 306}.  
\bibitem[(2008)] {Romero} Reynoso, M.M.; Romero, G.E.; Christiansen, H.R. Production of gamma rays and neutrinos in the dark jets of the microquasar SS433. {\em Mon. Not. R. Astron. Soc.} {\bf 2008}, {\em 387}, 1745--1754. 
\bibitem[(2009)] {Reynoso} Reynoso, M.M.; Romero, G.E. Magnetic field effects on neutrino production in microquasars. {\em Astron. Astrophys.} {\bf 2009}, {\em 493}, 1--11. 
\bibitem[(2006)] {Bosch-Ramon} Bosch-Ramon, V.; Romero, G.E.; Paredes, J.M. A broadband leptonic model for gamma-ray emitting microquasars. {\em Astron. Astrophys.} {\bf 2006}, {\em 447}, 263--276.
\bibitem[(1990)] {Begelman} Begelman, M.C.; Rudak, B.; Sikora, M. Consequences of Relativistic Proton Injection in Active Galactic Nuclei. {\em Astrophys. J.} {\bf 1990}, {\em 362}, 38.
\bibitem[(1999)] {Berezhko} Berezhko, E.G.; Ellison, D.C. A Simple Model of Nonlinear Diffusive Shock Acceleration. {\em Astrophys. J.} {\bf 1999}, {\em 526}, 385--399.
\bibitem[(1999)] {Gallant} Gallant, Y.A.; Achterberg, A. Ultra-high-energy cosmic ray acceleration by relativistic blast waves. {\em Mon. Not. R. Astron. Soc.} {\bf 1999}, {\em 305}, L6--L10.
\bibitem[(2000)] {Kirk} Kirk, J.G.; Guthmann, A.W.; Gallant, Y.A.; Achterberg, A. Particle Acceleration at Ultrarelativistic Shocks: An Eigenfunction Method. {\em Astrophys. J.} {\bf 2000}, {\em 542}, 235--242.
\bibitem[(2001)] {Achterberg} Achterberg, A.; Gallant, Y.A.; Kirk, J.G.; Guthmann, A.W. Particle acceleration by ultrarelativistic shocks: Theory and simulations. {\em Mon. Not. R. Astron. Soc.} {\bf 2001}, {\em 328}, 393--408.
\bibitem[(2003)] {Romero2003} Romero, G.E.; Torres, D.F.; Kaufman Bernadó, M.M.; Mirabel, I.F. Hadronic gamma-ray emission from windy microquasars. {\em Astron. Astrophys.} {\bf 2003}, {\em 410}, L1--L4.
\bibitem[(2007)] {Romero2007} Romero, G.E.; Okazaki, A.T.; Orellana, M.; Owocki, S.P. Accretion vs. colliding wind models for the gamma-ray binary LS I +61 303: An assessment. {\em Astron. Astrophys.} {\bf 2007}, {\em 474}, 15--22.
\bibitem[(1982)] {Friend} Friend, D.B.; Castor, J.I. Radiation driven winds in X-ray binaries. {\em Astrophys. J.} {\bf 1982}, {\em 261}, 293--300. 
\bibitem[(2006)] {Kelner2006} Kelner, S.R.; Aharonian, F.A.; Bugayov, V.V. Energy spectra of gamma rays, electrons, and neutrinos produced at proton-proton interactions in the very high energy regime. {\em Phys. Rev. D} {\bf 2006}, {\em 74}, 034018.
\bibitem[(2020)] {Papav-Papad-Kosm-JPCon} Papavasileiou, T.V.; Papadopoulos, D.A.; Kosmas, T.S. Astrophysical magnetohydrodynamical outflows in the extragalactic binary system LMC X-1. {\em J. Phys. Conf. Ser.} {\bf 2020}, {\em 1730}, 012138.
\bibitem[(2021)] {Papad-Papav-Kosm-JPCon} Papadopoulos, D.A.; Papavasileiou, Th.V.; Kosmas, T.S. High energy neutrino and gamma-ray emissions from the jets of M33 X-7 microquasar. {\em J. Phys. Conf. Ser.} {\bf 2021}, {\em 1730}, 012137.
\bibitem[(2017)] {Ahnen} Ahnen, M.L.; Ansoldi, S.; Antonelli, L.A.; Arcaro, C.; Babi´c, A.; Banerjee, B.; Bangale, P.; de Almeida, U.B.; Barrio, J.A.; González, J.B.; et al. Search for very high-energy gamma-ray emission from the microquasar Cygnus X-1 with the MAGIC telescopes. {\em Mon. Not. R. Astron. Soc.} {\bf 2017}, {\em 472}, 3474--3485.
\bibitem[(2017)] {Hyde} Hyde, E.A.; Russell, D.M.; Ritter, A.; Filipovi´c, M.D.; Kaper, L.; Grieve, K.; O’Brien, A.N. LMC X-1: A New Spectral Analysis of the O-star in the Binary and Surrounding Nebula. {\em Astron. Soc. Pac.} {\bf 2017}, {\em 129}, 094201.
\bibitem[(2016)] {Aartsen2016} Aartsen, M.G.; Abraham, K.; Ackermann, M.; Adams, J.; Aguilar, J.A.; Ahlers, M.; Ahrens, M.; Altmann, D.; Andeen, K.; Anderson, T.; et al. Searches for Sterile Neutrinos with the IceCube Detector. {\em Phys. Rev. Lett.} {\bf 2016}, {\em 117}, 071801.
\bibitem[(2016)] {Adrian-Martinez} Adrián-Martínez, S.; Ageron, M.; Aharonian, F.; Aiello, S.; Albert, A.; Ameli, F.; Anassontzis, E.; Andre, M.; Androulakis, G.; Anghinolfi, M.; et al. Letter of intent for KM3NeT 2.0. {\em J. Phys. G Nucl. Part Phys.} {\bf 2016}, {\em 43}, 084001.  
\bibitem[(2005)] {Cherepashchuk} Cherepashchuk, A.M.; Sunyaev, R.A.; Fabrika, S.N.; Postnov, K.A.; Molkov, S.V.; Barsukova, E.A.; Antokhina, E.A.; Irsmambetova, T.R.; Panchenko, I.E.; Seifina, E.V.; et al. INTEGRAL observations of SS433: Results of a coordinated campaign. {\em Astron. Astrophys.} {\bf 2005}, {\em 437}, 561--573.
\bibitem[(1970)] {Blumenthal} Blumenthal, G.R.; Gould, R.J. Bremsstrahlung, Synchrotron Radiation, and Compton Scattering of High-Energy Electrons Traversing Dilute Gases. {\em Rev. Mod. Phys.} {\bf 1970}, {\em 42}, 237--271.

\bibitem[(2007)] {Khangulyan} Khangulyan, D.; Hnatic, S.; Aharonian, F.; Bogovalov, S. TeV light curve of PSR B1259–63/SS2883. {\em Mon. Not. R. Astron. Soc.} {\bf 2007}, {\em 380}, 320--330.
\bibitem[(2011)] {Orosz2011} Orosz, J.A.; McClintock, J.E.; Aufdenberg, J.P.; Remillard, R.A.; Reid, M.J.; Narayan, R.; Gou, L. The mass of the black hole in cygnus X-1. {\em Astrophys. J.} {\bf 2011}, {\em 742}, 84.
\bibitem[(2009)] {Orosz2009} Orosz, J.A.; Steeghs, D.; McClintock, J.E.; Torres, M.A.P.; Bochkov, I.; Gou, L.; Narayan, R.; Blaschak, M.; Levine, A.M.; Remillard, R.A.; et al. A new dynamical model for the black hole binary lmc X-1. {\em Astrophys. J.} {\bf 2009}, {\em 697}, 573--591. 
\bibitem[(2001)] {Stirling} Stirling, A.; Spencer, R.; de La Force, C.; Garrett, M.; Fender, R.; Ogley, R. A relativistic jet from Cygnus X-1 in the low/hard X-ray state. {\em Mon. Not. R. Astron. Soc.} {\bf 2001}, {\em 327}, 1273--1278.
\bibitem[(2003)] {Atoyan} Atoyan, A.M.; Dermer, C.D. Neutral Beams from Blazar Jets. {\em Astrophys. J.} {\bf 2003}, {\em 586}, 79.  
\bibitem[(2008)] {Kelner} Kelner, S.R.; Aharonian, F.A. Energy spectra of gamma rays, electrons, and neutrinos produced at interactions of relativistic protons with low energy radiation {\em Phys. Rev. D} {\bf 2008}, {\em 78}, 034013.
\bibitem[(2011)] {Torres} Torres, D.F.; Reimer, A. Hadronic beam models for quasars and microquasars. {\em Astron. Astrophys.} {\bf 2011}, {\em 528}, L2. 
 
 

\end{thebibliography}



\end{document}